\documentclass{article}
\usepackage{spconf,amsmath,graphicx}
\usepackage{color}
\usepackage{multirow}
\usepackage{booktabs}
\usepackage{threeparttable}
\usepackage{graphicx}  
\usepackage{subcaption} 
\usepackage{adjustbox}
\usepackage{flushend}

\newcommand{\ie}{i.\,e.,\ }

\title{Enhanced Heart Murmur Detection via Branchformer with Uncertainty Estimation}
\title{Intelligent Cardiac Auscultation for Murmur Detection via Parallel-Attentive Models with Uncertainty Estimation}
\name{Zixing~Zhang$^1$, Tao~Pang$^1$*, Jing~Han$^2$*, Bj\"orn W.\ Schuller$^3$\thanks{$^*$ Corresponding authors. pang\_tao@hnu.edu.cn, jh2298@cam.ac.uk}}
\address{$^1$ College of Computer Science and Electronic Engineering, Hunan University, China\\
$^2$ Department of Computer Science and Technology, University of Cambridge, UK \\ 
$^3$ GLAM, Department of Computing, Imperial College London, UK}

\usepackage{multirow}
\begin{document}
\ninept
\maketitle
%
\begin{abstract}
Heart murmurs are a common manifestation of cardiovascular diseases and can provide crucial clues to early cardiac abnormalities. While most current research methods primarily focus on the accuracy of models, they often overlook other important aspects such as the interpretability of machine learning algorithms and the uncertainty of predictions. This paper introduces a heart murmur detection method based on a parallel-attentive model, which consists of two branches: One is based on a self-attention module and the other one 
is based on a convolutional network. Unlike traditional approaches, this structure is better equipped to handle long-term dependencies in sequential data, and thus effectively captures the local and global features of heart murmurs. Additionally, we acknowledge the significance of understanding the uncertainty of model predictions in the medical field for clinical decision-making. Therefore, we have incorporated an effective uncertainty estimation method based on Monte Carlo Dropout into our model. Furthermore, we have employed temperature scaling to calibrate the predictions of our probabilistic model, enhancing its reliability. In experiments conducted on the CirCor Digiscope dataset for heart murmur detection, our proposed method achieves a weighted accuracy of 79.8\,\% and an F1 of 65.1\,\%, representing state-of-the-art results.

\end{abstract}
%


\section{Introduction}
\vspace{-6pt}

Cardiovascular diseases (CVDs), \ie a group of disorders of the heart and blood vessels, are the major cause of death globally~\cite{balakumar2016prevalence}. 
Despite the rapid development of advanced diagnosis approaches, such as echocardiogram, cardiac computed tomography, and cardiac magnetic resonance imaging, the reduction of related mortality is still challenging. This is largely attributable to the requirements of high expenditure, frequent hospital visits, and professional physicians. Thus, cardiac auscultation (CA), a conventional, low-cost, non-invasive, and portable screening technique that can listen to heart sounds (\textit{aka} phonocardiogram [PCG]) and then evaluate cardiac function and identify any potential murmurs (abnormalities) or pathologies, still plays an indispensable role in CVDs' screening and diagnosis. Nevertheless, CA demands skilled professionals, the population of whom is, unfortunately, decreasing because of the long training process through theory and practical clinical experience~\cite{noor2013heart}. Therefore, developing an automated and intelligent decision system for CA by leveraging large-scale recordings is more needed than ever, aiding doctors in more accurately identifying and diagnosing heart conditions~\cite{chen2018heart}. 

For this reason, many studies have been committed to exploring machine learning approaches in this domain~\cite{ren2018learning, ren2023comprehensive}, 
including Support Vector Machines (SVMs), Convolutional Neural Networks (CNNs), Recurrent Neural Networks (RNNs) and their Long Short-Term Memory (LSTM) version. However, these models have recently and frequently shown limited capability in capturing complex spatiotemporal features both locally and globally, resulting in a suboptimal prediction performance for murmur detection from phonocardiogram recordings~\cite{ren2023comprehensive}. To address this issue, we introduce a \textit{parallel-attentive transformer} model that consists of two different branches: One contains a conventional multi-headed self-attention block, and the other one is structured by multilayer perceptions and convolutional networks. The former branch aims to capture global context information, whereas the latter one is mainly designed to learn more local context information. By merging the information flows from the two branches, this model shows more efficiency in capturing the salient representation for murmur detection than other classic models. This model was recently introduced for speech recognition and understanding, and achieved state-of-the-art (SOTA) results~\cite{peng2022branchformer}, but was never investigated in such low-frequency audio signals. 

Moreover and notably, most of the previous studies, if not all, ignored how to ignite the reliability of the model predictions. A hard model prediction can hardly persuade cardiologists or patients to trust the decisions from these automatic systems~\cite{lahav2018interpretable}. The lack of a metric to evaluate the reliability of model predictions greatly hinders their deployment in realistic application scenarios. To close this gap, we further introduce an \textit{uncertainty} strategy with Monte Carlo Dropout~\cite{gal2016dropout} into the model predictions, turning the hard decisions into soft ones. The uncertainty value estimates the confidence of the approach's predictions, which simulate the human decision-making process based on their knowledge and skills. Furthermore, to calibrate the uncertainty estimation, we apply a post-process of temperature scaling~\cite{guo2017calibration} consecutively, which can help adjust the prediction confidence to better reflect the true likelihood of each class. 

\begin{figure*}[!h]
   \centering
    \includegraphics[width=\textwidth]{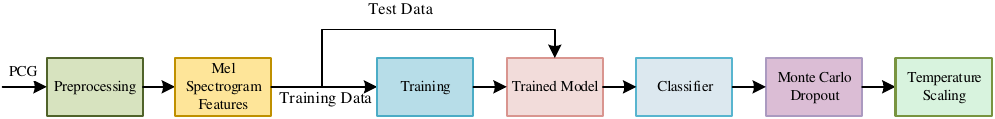}
    \caption{The flowchart of the proposed murmur detection system.
    }
    \vspace{-10pt}
    \label{fig:overview} 
\end{figure*}

In short, the main contributions of this paper can be summarised in the following three aspects: 
\begin{itemize}
    \item We propose an innovative cardiac murmur detection method based on a parallel-attentive transformer model, which holds the potential to extract more salient spatiotemporal representations locally and globally than other models.  
    \item We introduce an uncertainty estimation, \ie Monte Carlo Dropout, along with the murmur detection process, and a temperature scaling post-process to compensate for the estimation. This considerably promotes the reliability of the model prediction for murmur detection. 
    \item We evaluate the introduced murmur detection system in a publicly available and widely used large-scale dataset, \ie the CirCor Digiscope dataset~\cite{oliveira2021circor}, and achieve a new SOTA performance. 
\end{itemize}

\vspace{-7pt}
\section{Related Work}
\vspace{-5pt}
Prior to conducting this study, a substantial body of related research has explored the application of machine learning in the field of heart sound classification. Many researchers have focused on feature extraction and classifier development for heart sound classification. 
Abduh et al.~\cite{abduh2020classification} employed Mel-frequency cepstral coefficients (MFCCs) based on the Short-Time Fourier Transform (STFT) and employed various classifiers, including SVMs, K-Nearest Neighbors, and ensemble classifiers
for heart sound classification. In~\cite{ghosh2019automated}, the authors propose to use the wavelet synchronisation transformation to obtain a time-frequency matrix extracting magnitude and phase information from the segmentation period of the PCG signal and to classify the signal using a random forest classifier. 
In recent years, with the development of deep learning, some researchers have begun to explore its application in the field of heart disease diagnosis.
In~\cite{chen2016automatic}, the STFT magnitude spectrum was employed as input for a convolutional neural network in heart sound classification.
In~\cite{elola2023beyond}, researchers fed the Mel spectrogram features of each PCG record into an ensemble of 15 convolutional residual neural networks. This ensemble was used to classify heart murmurs for each patient, categorising the murmurs into three classes: ``absent'', ``soft'', and ``loud''.
Apart from utilising CNNs for classifying heart sounds, RNNs and LSTMs were also employed in heart sound classification. For instance, in~\cite{alkhodari2021convolutional}, the authors combined CNNs with bidirectional LSTM to automatically extract features related to the temporal and spatial domains of the signal. 
\vspace{-6pt}
\section{Methodology}
\vspace{-5pt}
The process diagram of the proposed method is illustrated in Fig.~\ref{fig:overview}. Firstly, data is preprocessed, and then, Mel spectrogram features are extracted. Model learning and parameter optimisation are conducted on the training dataset. During the training process, the model performance is monitored using the validation dataset, and the best-performing model is saved based on validation results. Finally, the ultimate classification results are obtained on the test dataset. To comprehensively assess the model's performance, the Monte Carlo Dropout method is employed to estimate the model's uncertainty. Subsequently, the temperature scaling method is applied to the validation dataset to adjust the temperature parameter, calibrating the model's output probabilities. Lastly, probability calibration is performed on the test dataset to make the model's outputs more reliable.

\vspace{-5pt}
\subsection{Parallel-Attentive Modelling}
\begin{figure}[t]
    \centering
    \vspace{-5pt}
    \includegraphics[width=\linewidth, trim=360 110 290 90,clip]{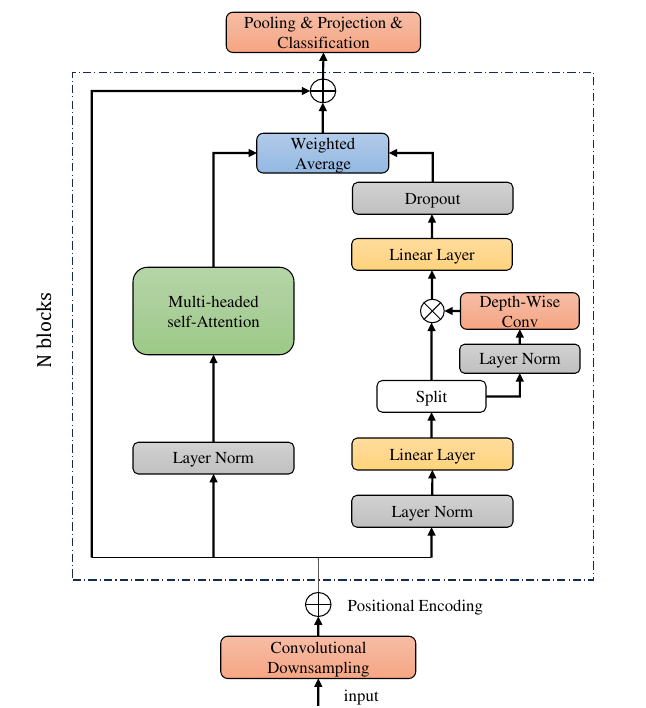}
    \caption{Illustration of the Parallel-Attentive Modelling. 
    }
    \label{fig:branchformer}
\end{figure}

\vspace{-5pt}
The detailed structure of the used parallel-attentive modelling is shown in Fig.~\ref{fig:branchformer}. It employs a parallel structure comprising self-attention modules and convolutional modules to better capture crucial information within sequences. To reduce computational complexity while retaining vital details, we introduce a convolutional subsampling module for time-based downsampling of feature sequences, downsampling the input by a factor of four. The downsampled results are then combined with positional encoding and fed into two branches. One branch, after LayerNorm, utilises a multi-head self-attention mechanism with relative positional encoding. The primary role of this branch is to acquire longer-range contextual information within the sequence, facilitating the comprehension of global relationships. The other branch is centred around the use of depth-wise convolution, aimed at capturing local information. This architecture effectively amalgamates both global and local information, leading to improved model performance. 

\vspace{-5pt}
\subsection{Uncertainty Estimation with Monte Carlo Dropout}
\vspace{-5pt}
Model uncertainty can help doctors and clinical experts understand the prediction confidence of the models, allowing them to make better medical decisions. If the model exhibits a high level of uncertainty for a particular prediction, healthcare professionals may exercise greater caution in considering whether to proceed with further diagnostic or treatment steps. In classification problems, the Softmax function transforms the scores for each category into probabilities, with the probabilities summing to 1. This can sometimes lead to situations where the model's output is incorrect despite having a high probability value, as the model may output an incorrect value with high confidence. 
To overcome this problem, research tasks have proposed many methods, such as Bayesian Neural Networks \cite{teye2018bayesian}, Deep ensembles \cite{lakshminarayanan2017simple}, and Monte Carlo Dropout methods. When the uncertainty of the model is obtained by Monte Carlo Dropout, the dropout layer is turned on during inference and the forward propagation process is performed $N$ times for the same input. The probability of $N$ times prediction is
\vspace{-8pt}
\begin{equation}
p(y=c|x)=\frac{1}{N}\sum_{n=1}^{N} \text{softmax}(f^{\hat{W}_n}(x)), 
\end{equation}
where $\hat{W}_t$ denotes the uncertainty parameter of the model when the dropout layer is turned on.
The uncertainty of the model is measured using the following Shannon entropy:
\vspace{-8pt}
\begin{equation}
H(p)=-\sum_{k=1}^{K}p_K \text{log}p_K,
\end{equation}
where $K$ is the number of categories. 
\vspace{-5pt}
\subsection{Calibration with Temperature Scaling}
\vspace{-5pt}
Temperature scaling is a parametric calibration method that introduces a temperature coefficient to adjust the output logits of the model, effectively amplifying or dampening them. This adjustment is made to increase or decrease the distances between the predicted scores for different categories. For each input, denoted as \(x_i\), the neural network generates a class prediction value $\hat{y}_i$ along with a confidence score $\hat{p}_i$. The vector \(z_i\) represents the output logits. Thus, the probability of each class can be expressed by
\vspace{-8pt}
\begin{equation}
\text{softmax}(z_i)^{(k)}=\frac { \text{exp}(z_i^{(k)}) }{ \sum_{j=1}^{K} \text{exp}(z_i^{(k)}) }.
\end{equation}
Before applying temperature scaling, the confidence score 
$\hat{p}_i$ is represented as
\vspace{-8pt}
\begin{equation}
    \hat{p}_i=\underset{\text{k}}{\text{max}} \:\text{softmax}(z_i)^{(k)}.
\end{equation}
Temperature scaling involves scaling the logits within the formula 
\vspace{-5pt}
\begin{equation}
    \text{softmax}(z_i)^{(k)}=\frac { \text{exp}(z_i^{(k)}/T)}{ \sum_{j=1}^{K} \text{exp}(z_i^{(k)}/T) }, 
\end{equation}
where $T$ is a learnt temperature scaling parameter, which is optimised on the validation set to align the predicted probabilities with the actual uncertainty of the validation set. This process helps to improve the calibration of the model's confidence estimation.

\vspace{-10pt}
\section{Experiments}
\vspace{-5pt}
\subsection{Selected Dataset}
\vspace{-5pt}
In this study, we performed murmur detection using the CirCor Digiscope dataset \cite{oliveira2021circor} and classified the murmurs into the following three categories: presence of murmur, absence of murmur, and unknown (presence of noisy situations that the annotator could not determine).

The CirCor DigiScope Dataset is currently the largest dataset of heart sounds in children and consists of 5,272 recordings from 1,568 patients, sampled at 4,000\,Hz. In the majority of patients, the recordings cover four main auscultation positions of the aortic valve (AV), the pulmonary valve (PV), mitral valves (MV), and tricuspid valves (TV), with some recordings taken elsewhere and labelled as Phc. However, individual patients' recordings may cover fewer than four positions; also, very few patients had multiple recordings in the same position. Details of the dataset are shown in Table~\ref{tab:dataset}.

\begin{table}
\centering
\caption{The statistical patient and recording (in bracket) distribution of the CirCor Digiscope dataset. 
}
\label{tab:dataset}
\begin{tabular}{crrrr }
\toprule
 & \#absent & \#present & \#unknown & $\sum$\\
\midrule
train & 695\,(2\,391) & 179\,(616) & 68\,(156) & 942\,(3\,163)\\
val. & 106\,(357) & 27\,(92)& 16\,(37) & 149\,(486)\\
test & 343\,(1\,186)& 99\,(357) & 35\,(85) & 477\,(1\,623)\\
\midrule
$\sum$ &1\,141\,(3\,929) &305\,(1\,065) &119\,(278) &1\,568\,(5\,272) \\ 
\bottomrule
\end{tabular}
\vspace{-8pt}
\end{table}

\vspace{-2pt}
\subsection{Implementation Details}
\vspace{-3pt}
\textbf{Data preprocessing}: 
Each patient's PCG recording is segmented using a 3-second sliding window with a 1-second overlap, and a 25\,ms frame length and 12.5\,ms step size are used to compute the Mel spectrogram with an FFT size of 512 and the number of Mel frequency bins of 128. This results in a 128\,x\,241 feature map for each data segment.

\vspace{.2cm}
\noindent\textbf{Training Setup}: 
Our model uses 6 parallel-attentive layers, each with 4 attention heads, and each attention head has a dimension of 128. Dropout rate is 0.1. Further, due to extreme data imbalance, a weighted cross-entropy loss is used to train the model with weight ratios of 1:5:3 for absent:present:unknown classes. The learning rate is 1e-4 as an initial value, and the Adam optimiser is employed with a weight decay. Each mini-batch has a size of 128 samples. During the training process, if the loss on the validation set does not decrease for five consecutive epochs, the learning rate is reduced by half. The total number of epochs is set to 30, and the best-performing model on the validation set is saved during training.

\vspace{.2cm}
\noindent\textbf{Determine patient-level predictions}: 
 As the ultimate goal of the research is to classify murmurs at a patient level, it is necessary to first determine the predictions for each patient record, and then derive a patient prediction from the record predictions.

For the record classification, the prediction for a record is determined by the majority predictions of its segments, with a priority given to `present', `unknown', and `absent'.

For the patient classification, if any of the record segments are predicted as `present', the patient is determined as `present'. Otherwise, if any of the record segments are classified as `unknown', the patient is identified as `unknown'. If none of the record segments are recognised as `present' or `unknown', it indicates that all segments are `absent' and the patient is determined as `absent'.

\vspace{.2cm}
\noindent\textbf{Performance Metrics}:
In order to evaluate the method we employed, the experiments utilise weighted accuracy and F1-score as evaluation metrics, with weighted accuracy being introduced by The George B.\ Moody PhysioNet Challenge 2022~\cite{reyna2022heart}. Weighted accuracy assigns greater importance to the ``present'' and ``unknown'' categories, helping to reduce missed diagnoses and improve the reliability of early disease detection.

(1) weighted accuracy: The definition of weighted accuracy is as follows:
\vspace{-8pt}
\begin{equation}
\text{Acc}_w=\frac{m_{AA}+5m_{PP}+3m_{UU}}{
\sum_{i}m_{iA}+5\sum_{i}m_{iP}+3\sum_{i}m_{iU}
}. 
\end{equation}
The confusion matrix for this weighted accuracy is shown in Table~\ref{tab:Confusion_matrix_template}.

\begin{table}[!t]
\centering
\caption{Confusion matrix used for calculating the weighted accuracy.}
\vspace{-8pt}
\label{tab:Confusion_matrix_template}
\begin{tabular}{c |c| c c c }
\toprule
& &\multicolumn {3}{c}{prediction}\\
\cmidrule{3-5}
&   & absent & present & unknown\\
\midrule
 \multirow{3}{3em}{ground truth} & absent & \(m_{AA}\) & \(m_{PA}\) & \(m_{UA}\) \\

 & present & \(m_{AP}\) & \(m_{PP}\) & \(m_{UP}\) \\
 & unknown & \(m_{AU}\) & \(m_{PU}\) & \(m_{UU}\) \\
\bottomrule
\end{tabular}
\end{table}


(2) Expected Calibration Error (ECE)~\cite{naeini2015obtaining}: ECE is a metric used to assess the calibration performance of classification models, measuring the degree of consistency between the model's probability of predictions and the actual observations. The calculation method for ECE is as follows:
\vspace{-10pt}
\begin{equation}
    \text{ECE}=\sum_{i=1}^{M}\frac {|B_i|}{ n }|\text{accuracy}(B_i)-\text{confidence}(B_i)|,
\end{equation}
where $M$ is the number of intervals (M = 15 in our experiments), and $B_i$ denotes each interval.

\vspace{-10pt}
\section{Results and Discussion}
\vspace{-5pt}
\subsection{Accuracy}
\vspace{-5pt}
The results of the proposed model are shown in Table~\ref{tab:result}. When compared with the top three performers in The George B.\ Moody PhysioNet Challenge 2022, our method beats all of them and achieves the best performance. The weighted accuracy of our proposed model for patient-level heart murmur classification is 79.8\%, indicating that our model can better identify and classify heart murmurs. Additionally, our model has made an obvious improvement in terms of F1-score. 
In summary, our research results demonstrate that the use of parallel attention mechanisms can significantly improve the performance of heart murmur classification tasks ($p<.05$ with a one-tailed $z$-test), which may largely be thanks to its efficiency in capturing local and global information simultaneously. 

\begin{table}[!t]
\centering
\caption{Performance comparison in terms of weighted accuracy and macro-F1 between the proposed modelling and other SOTA approaches. 
}
\label{tab:result}
\vspace{-8pt}
\begin{tabular}{c c c }
\toprule
 models & Weighted Accuracy [\%] &macro-F1[\%]  \\
\midrule
CNN~\cite{lu2022lightweight} & 78.0&61.9 \\
RNN-HSMM~\cite{mcdonald2022detection} & 77.6&62.3 \\
HMS-Net~\cite{xu2022hierarchical} & 77.6&64.7 \\
Proposed & \textbf{79.8} &\textbf{65.1} \\
\bottomrule
\end{tabular}
\end{table}

\begin{table}[!t]
\centering
\caption{The ECE of the murmur detection models in the segment and patient levels before and after the probability calibration.}
\label{tab:ECE}
\vspace{-8pt}
\begin{tabular}{c c c }
\toprule
& uncalibrate & calibrated \\
 \midrule
segment & 0.063 & 0.049\\
patient & 0.098 & 0.043 \\
\bottomrule
\end{tabular}
\vspace{-10pt}
\end{table}

\vspace{-8pt}
\subsection{Uncertainty and Calibration}
\vspace{-5pt}
We performed 30 times of random forward propagation and averaged the results. Subsequently, we calibrated the model's output and calculated the ECE.
As the primary objective of this study is patient-level classification, our focus is on extending the model's uncertainty estimation from the segment level to the entire patient level to better serve our research goals. We employ the following methods to accomplish this target. Starting with the mapping ``from segment to record'', we first determine the predictions for records and then take the average of segments that align with the record's prediction. Subsequently, with the mapping ``from record to patient'', we initially ascertain the patient's predictions and then average records that match the patient's predictions.
Table~\ref{tab:ECE} presents the ECEs of the uncertainty estimation before and after applying the calibration with temperature scaling. Obviously, the ECEs calculated on either segments or patients are remarkably reduced after the probability calibration.

\begin{figure}[!t]
\begin{subfigure}{0.25\textwidth}
\centering
\includegraphics[width=0.8\linewidth, height=3cm]{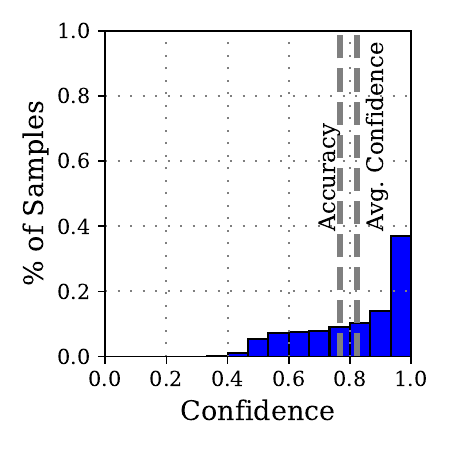}
\end{subfigure}
\vspace{-6pt}
\begin{subfigure}{0.25\textwidth}
\includegraphics[width=0.8\linewidth, height=3cm]{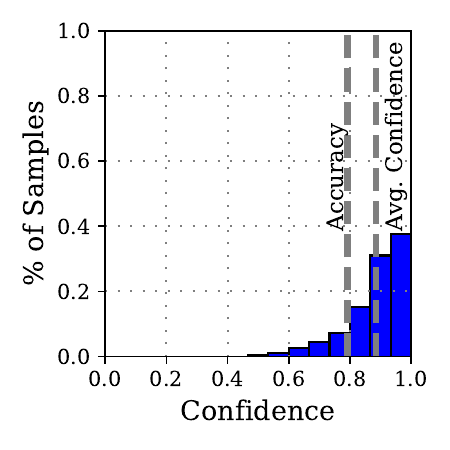}
\end{subfigure}
\begin{subfigure}{0.25\textwidth}
\centering
\includegraphics[width=0.8\linewidth, height=3cm]{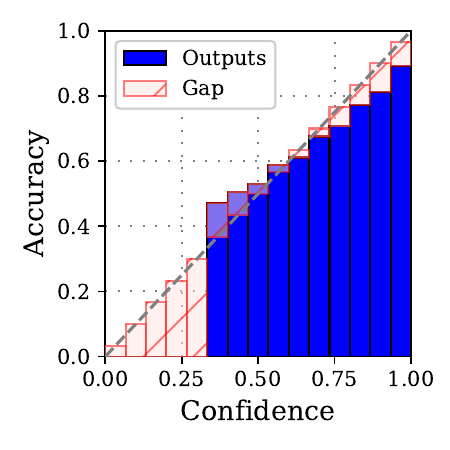} 
\end{subfigure}
\begin{subfigure}{0.25\textwidth}
\includegraphics[width=0.8\linewidth, height=3cm]{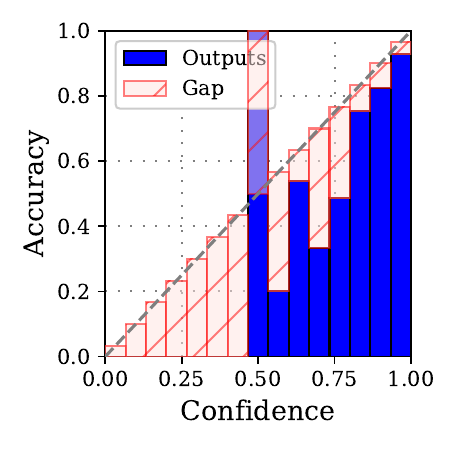}
\end{subfigure}
\vspace{-20pt}
\caption{Confidence Histogram (above) and Reliability Diagram (below) on the segments (left) and patients (right) \textbf{before} the confidence calibration.}
\label{fig:uncalibrate}
\end{figure}

\begin{figure}[!t]
\begin{subfigure}{0.25\textwidth}
\centering
\includegraphics[width=0.8\linewidth, height=3cm]{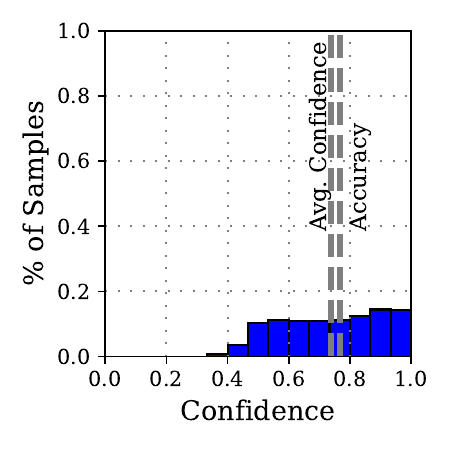} 
\end{subfigure}
\vspace{-6pt}
\begin{subfigure}{0.25\textwidth}
\includegraphics[width=0.8\linewidth, height=3cm]{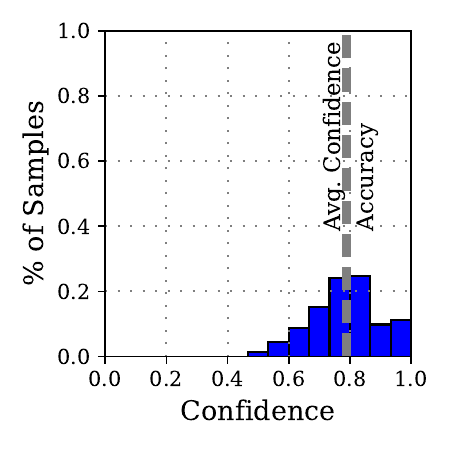}
\end{subfigure}
\begin{subfigure}{0.25\textwidth}
\centering
\includegraphics[width=0.8\linewidth, height=3cm]{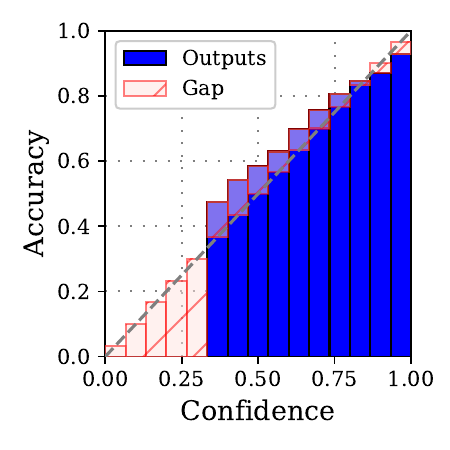} 
\end{subfigure}
\begin{subfigure}{0.25\textwidth}
\includegraphics[width=0.8\linewidth, height=3cm]{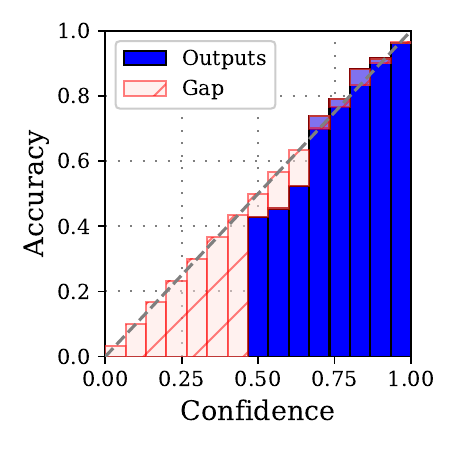}
\end{subfigure}
\vspace{-20pt}
\caption{Confidence Histogram (above) and Reliability Diagram (below) on the segments (left) and patients (right) \textbf{after} the confidence calibration.}
\vspace{-10pt}
\label{fig:calibrated}
\end{figure}
Additionally, we have visualised representations in the form of graphs to illustrate prediction probabilities (confidence) and reliability diagrams before and after calibration (see Fig.~\ref{fig:uncalibrate} and Fig.~\ref{fig:calibrated}). By examining these charts, we can see that with this calibration process, the prediction confidence of models aligns more closely with the accuracy, which suggests the effectiveness of our calibration method.

\vspace{-8pt}
\section{Conclusion}
In this paper, we presented a parallel-attentive modelling for cardiac murmur detection. The leading-board results imply the introduced model can efficiently capture local and global context information for such sequential heart sound signals, 
providing doctors with a more accurate diagnostic tool. Moreover, the introduction of techniques such as Monte Carlo methods and temperature scaling further enhanced the model's reliability, yielding more reliable predictions in uncertainty assessment and calibration. The application of these methods provides strong support for the accuracy and clinical applicability of cardiac murmur detection. It is anticipated that this achievement will propel the automated murmur detection system closer to being implemented in clinical practice in future. 
\vfill\pagebreak
\bibliographystyle{IEEEbib}
\bibliography{refs}

\end{document}